# Intrinsic local symmetry-breaking in nominally cubic paraelectric BaTiO$_3$

Xin-Gang Zhao,[1] Oleksandr I. Malyi,[1] Simon J.L. Billinge,[2] and Alex Zunger[1,*]

1. Renewable and Sustainable Energy Institute, University of Colorado, Boulder, Colorado 80309

2. Department of Applied Physics and Applied Mathematics, Fu Foundation School of Engineering and Applied Sciences, Columbia University, New York, NY 10027

*Corresponding author email: Alex.Zunger@colorado.edu

Whereas low-temperature ferroelectrics have a well understood ordered spatial dipole arrangement, the fate of these dipoles in paraelectric phases remains poorly understood. Using density functional theory (DFT), we find that unlike the case in conventional non-polar ABO$_3$ compounds illustrated here for cubic BaZrO$_3$, the origin of the distribution of the B site off-centering in cubic paraelectric such as BaTiO$_3$ is an intrinsic, energy stabilizing symmetry breaking. Minimizing the internal energy E of a constrained cubic phase already reveals the formation of a distribution of intrinsic local displacements that (i) mimic the symmetries of the low temperature phases, while (ii) being the precursors of what finite temperature DFT Molecular Dynamics finds as thermal motifs. The implications of such symmetry breaking on the microscopic structures and anomalous properties in these kinds of PE materials are discussed.

Oxide perovskites $ABO_3$, typified by $BaTiO_3$ (BTO), have local dipoles $\{\mu^{(i)}_{local}\}$, formed by polar atomic displacements off high symmetry positions $i$. At low temperatures, these can organize into ferroelectric (FE) long-range-ordered (LRO) structures. In contrast, above the Curie temperature $T_C$ ~401 K, BTO transforms into a cubic paraelectric (PE) phase. [1] Crystallographically, this high-temperature BTO structure—as well as experimentally reported ~90 other oxides $ABO_3$ [2,3] and scores of halide [4,5] perovskites are classified as Pm-3m space group symmetry. A central question in this field [6–16] regards the nature of the spatial configurations of the local $\{\mu^{(i)}_{local}\}$ dipoles in the paraelectric phase. Suppose the Pm-3m paraelectric phase, consisting of a single, undeformed octahedral motif as its repeat unit (the monomorphous configuration), is taken literally rather than as a 'virtual crystal' average over configurations. In this case, the PE phase will not only have vanishing global dipole moment $\mu_{global}$ ~0, but also vanishing local dipoles on each site, $\{\mu^{(i)}_{local} = 0\}$ (a 'non-electric' configuration; Fig. 1a insert). Experimentally, in contrast, the pioneering 1968 diffuse scattering XRD measurements of Comes et al. [9] on the PE phase were interpreted as being due to the presence of nonzero local dipoles $\{\mu^{(i)}_{local}\}$ that persist locally in the high-temperature cubic phase of BTO. Other measurements also seem to indicate the off-centering of the Ti ions not only in the low temperature FE phases but also in the PE phase. This includes Nuclear Magnetic Resonance (NMR), [10] atomic pair distribution function (PDF) analysis of neutron powder diffraction data, [11–13] X-ray absorption fine structure (EXAFS), [14] as well as the observation of birefringence, [17] second harmonic generation (SHG), [18] and piezoelectricity in cubic oxide paraelectrics. [19,20] The SHG and piezoelectricity effects, however, are forbidden in the centrosymmetric nominal cubic phase, suggesting that locally the symmetry in PE BTO is lower. Although some sightings of such 'forbidden symmetries' appear to involve *extrinsic* factors (such as defects [21] or growth-induced non-ideal microstructures [22]), there are cases, most prominently BTO, [23] where intrinsic factors of symmetry-breaking appear to be at play. Other experiments, most notably inelastic neutron scattering, [24,25] suggest that the local dipoles appear only on cooling through the ferroelectric phase transition. The resulting controversy has held the perovskite community in fascination for over 50 years. This can be posed in the following way: Is the paraelectric condition of $\mu_{global}$ ~0 realized by having (a) $\{\mu^{(i)}_{local} = 0\}$ (symmetry unbroken, absence of local dipoles, as illustrated in Fig. 1a insert) or (b) via distribution of mutually compensating nonzero local dipoles $\{\mu^{(i)}_{local} \neq 0\}$ (Fig. 1d insert). What makes the resolution of this conundrum important is that the choice between (a) vs. (b) strongly affects the nature of predictions of calculations of spectroscopic and transport properties. For example, the use of the monomorphous view (a) ubiquitous in standard databases [26–29] as input to calculations often leads to a significant underestimation of band gaps. [30–33] On the other hand, possibility (b) has been generally interpreted by assigning local dipoles in the PE phase to *thermal fluctuations,* where atoms are oscillating thermally around their Wyckoff positions and time averaging to a zero polarization. However, such thermal effects do not account for the intrinsic nature of the polarization implied by the macroscopic experiments. [34] Indeed, in principle, there is also a possible contribution from modes driven by intrinsic chemical bonding effects even before thermal motion is considered. These are known in inorganic compounds, such as degeneracy removal exemplified by the Jahn-Teller distortions of *d* orbital impurities in insulators, [35] or by the lone-pair *s* orbitals of Sn, Pb, or Bi centers in inorganic compounds, [36] or via sterically-induced octahedral rotations and tilting in perovskites, [37] all understood as a-thermal energy lowering symmetry-breaking.

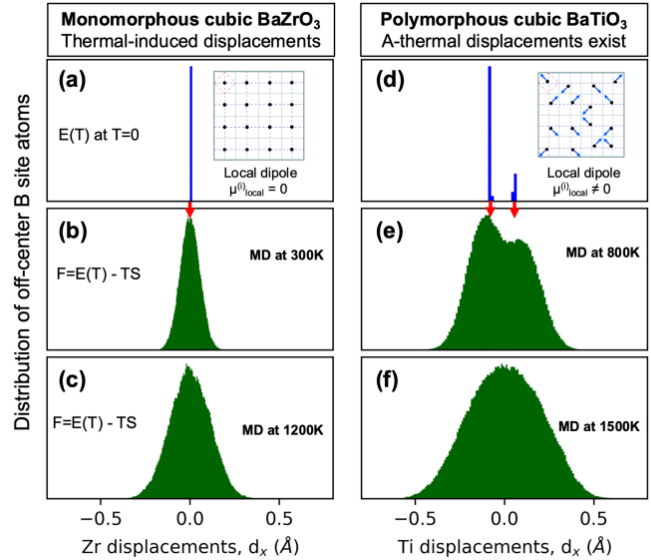

**Figure 1.** Distribution of projected B site displacements $d_x$ (compare Fig. S-1 for a 2D projection $(\theta, r)$) in ABO$_3$ with respect to the center of mass of an octahedron is shown for the cubic BaZrO$_3$ in (a-c), and for the intrinsically polymorphous cubic BaTiO$_3$ (in d-e). Inserts in (a, d) are the schematic illustrations of the B atomic displacements in models of cubic structure nominal (monomorphous) and polymorphous cubic. The displacements were taken from ~20,000 snapshots on equilibrated MD trajectories with *NPT* ensemble (*N*=64 fu/cell; *P*= 1 atm; timestep= 1 fs). The a-symmetry in (e) with respect to Ti displacements changed by less than 10% for simulation times of 20, 40, 60 ps.

The present work offers a different view on the *origin and properties* of the microscopic structure of paraelectric phases, using as an example the classic cubic paraelectric (PE) BaTiO$_3$. We start with first-principles energy minimization of the *internal energy* E, followed by a finite temperature molecular dynamics (MD) simulation of its free *energy F=E-TS.* We find that the displacement pattern in the finite temperature PE phase evident in MD follows a 'blueprint' encoded already by the a-thermal symmetry breaking that emerges from a constrained minimization of E for a cubic phase. The resulting 'polymorphous network' exhibits a pattern of intrinsic, a-thermal off-center displacement peaks (shown by the blue lines in Fig. 1d and the leading red arrows). This behavior is distinct from that in conventional 'monomorphous' compounds such as GaAs, Si, or BaZrO$_3$ (the latter shown in Fig 1(a, b, c)) that are not stabilized by intrinsic symmetry breaking. As temperature rises (Fig. 1e at 800 K), the underlying polymorphous pattern in BaTiO3 is initially retained at intermediate temperatures before it is overwhelmed by strong thermal motions at much higher temperatures (Fig. 1f, T=1500K). The latter displacement pattern is now centered symmetrically at the zero-displaced position, a behavior characteristic of conventional compounds at *all temperatures.* Our *ab-initio* calculated displacement pattern is consistent with the measured PDF [11–13]. The intrinsic symmetry breaking is enabled only if one allows a larger than minimal high symmetry (Pm-3m) unit cell which has the geometric flexibility to break symmetry--should the internal energy *E* be lowered.

This is a fundamentally important result with broad implications. It is an example where allowing the structure to have symmetry lowering distortions, even whilst respecting the global, 'average' cubic symmetry, that DFT can find *intrinsic symmetry broken structures that lower the energy of the system*. They are intrinsic in the sense that the distortions are not a result of extrinsic imperfections such as defects or doping. We believe that not all structures, but many structures will do this, especially ones with high susceptibilities to various applied forces such as the ferroelectrics discussed here. It not only explains poorly understood experimental observations in BTO but more broadly gives us a recipe for finding new functional materials by using DFT to search for previously unknown polymorphous network materials.

The present approach to the microscopic understanding of para phases allows us to learn a great deal about symmetry breaking. For example, upon projecting the displacement field on the irreducible representation, we find the *hidden polar FE-like mode* $\Gamma_4^-$, forbidden in the nominal cubic structure but consistent with the experimentally observed SHG [18] and piezoelectricity [19,20] that are disallowed in centrosymmetric ( e.g., Pm-3m) phases. It is significant that the molecular dynamics study using the classic force field (lacking explicit electronic degrees of freedom) demonstrate (Pasciak et al. [36]) *at* all temperatures a symmetric displacement field, producing zero net polarization and a centrosymmetric time-averaged structure, missing the observation of SHG and piezoelectricity. Furthermore, in contrast to commonly used approximations [38] that describe PE phases as an *averaged* high-symmetry cubic structures that are *non*-electric (vanishing global polarization because all local dipoles are also zero), we deduce that both structural properties and the ensuing electronic structure of PE phases cannot be described by such monomorphous cubic Pm-3m structure (used in 1000s of previous studies), nor as a long-range-ordered cubic AFE phase with single type of AFE modes associated with the single parabolic potential energy surface model commonly invoked to describe PE phases.

*Theoretical approach:* To understand the possible significance of symmetry-breaking within the PE phase and whether it emerges from intrinsic or thermal effects, we use a polymorphous DFT (polyDFT) approach. [39,40] In brief, it (i) does not restrict the *local* symmetry of the PE phase to that of a monomorphous primitive structure having a single local motif, (ii) does not use a long-range-ordered dipole model for the PE phase, [41] and (iii) explores both intrinsic (via minimization of the internal energy *E*) and dynamic (via thermal evolution of the free energy *F=E-TS* in first-principles molecular dynamics). To allow unimpeded local symmetry-breaking whilst respecting the *global* cubic symmetry, we constrain the lattice vectors to the macroscopically observed cubic shape but relax the internal energy *E* by allowing atoms to relax off high symmetry positions to minimize quantum forces on them. We use enlarged unit cells (here, 64 formula units (fu) containing 320 atoms; the results are stable against further cell enlargement). Supplementary Table S-1 demonstrates the rather weak dependence of the result on the initial configuration and on the random atomic nudges used to initialize the minimization. In this approach, the nature of the microscopic motifs in the para phase is determined by predictive energy minimization in a large supercell, rather than by modeling it at the outset as a single [41] or a couple of ordered structures. The results are analyzed to study the correlation between different displacements by projecting the displacement field on irreducible representations as did in Duyker et al. [42] We do so via first-principles DFT that retains explicit electronic and lattice degrees of freedom, unlike the fitted force fields that renormalize electronic degrees of freedom.[16] Whereas using a first-principles description limits supercell size and simulation time, the appearance of an active irreducible representation mode $\Gamma_4^-$ shown in Fig. 3 at all temperatures clearly predicts correlated disorder consistent with net polarization ~0.02 Å at 800 K and ~0.01 Å at 1200 K. However, it does not capture the few nm-scale correlated displacements that have been reported experimentally near the Curie temperature. [9]

*The distribution of Ti-O atomic pairs in paraelectric phase (Fig. 2):* We consider the partial radial distribution function (PRDF), $g(r)_{ab}$, between species *a* and species *b*, defined as

$$g(r)_{ab} = \frac{1}{N_a}\sum_{i=1}^{N_a} \frac{\sum_{j=1}^{N_b}\langle\delta(|r_{ij}-r|)\rangle}{4\pi r^2 dr} \quad (1)$$

where $N_a$ and $N_b$ refer to the total number of atoms for each species and $r_{ij}$ is the length of the vector from atom *i* to atom *j*. The PRDF of Ti-O pairs in the ground ferroelectric rhombohedral structure is shown in Fig. 2a and compared with the results of the polymorphous PE cubic network with intrinsic distortion (blue lines Fig. 2b) and with snapshots at 800 and 1200 K of the MD profile (Fig. 2c, d). Significantly, we find, in the polymorphous cubic PE phase, a splitting of the nearest-neighbor Ti-O peak (Fig. 2e). This splitting is due to the emergence of the set of local Ti off-center displacements (Fig. 2a). No constraint is placed on the directions of symmetry lowering displacements in the DFT calculations, which are guided simply by lowering the internal energy. However, we note that the displacements occur predominantly along {111} directions, which is consistent with the ground-state rhombohedral structure and the experimental observations of local structural probes, [9,11,14] resulting in the formation of three short (1.88 Å) and three long (2.13 Å) Ti-O bonds. A remnant of the nearest-neighbor Ti-O bond splitting is also observed in PRDFs obtained from our finite temperature MD calculations at temperatures of 800 and 1200 K (green lines shown in Fig. 2c, d). Interestingly, the polymorphous cubic PE phase also has local structural motifs that are not present in either the nominal cubic monomorphous approximant (Fig. 2e) or the low-temperature ferroelectric phase (Fig. 2a). The PRDF of the Ba-Ba, Ti-Ti, Ba-O, and O-O pairs in the polymorphous network (Fig. S-2) show relatively small broadening in the range of 1-8 Å. This might be because of the Goldschmidt tolerance factor [43] $t_{eff} = \frac{R_A+R_O}{\sqrt{2}(R_B+R_O)}$ of cubic BTO (1.06) is slightly larger than one, so octahedral tilting/rotation or A site displacements are small.

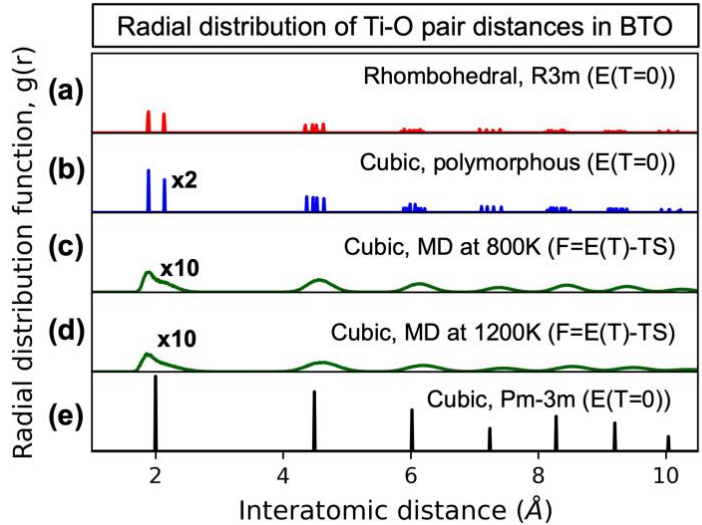

**Figure 2.** (a-e) Ti-O radial pair distribution functions (RPDF) before and after temperature set in. Here, (a) is the RPDF for the intrinsic rhombohedral phase (1fu/cell, red line), (b) is for the intrinsic polymorphous network (blue line), and (c) and (d) are from 100 snapshots equilibrium trajectories of MD simulation in the NPT ensemble at 1 atm, 800 and 1200 K with interval 0.1 ps. Finally, (e) shows RPDF in nominal cubic phase (1 fu/cell, black line).

*Internal energy-driven local symmetry breaking:* The characteristic of a polymorphous [39,40] phase of a material is that the total energy per atom is seen to initially decrease as the supercell size increases before saturating at a certain supercell size. We find that **cubic** BTO is polymorphous (average energy reduction -16 meV per formula unit relative to the monomorphous approximation), with predominantly Ti off-center displacements (Fig. 2b). Not all materials are polymorphous, as shown in Fig. 1a, the non-paraelectric $BaZrO_3$ has no symmetry breaking except for the thermal motion. Indeed, we find that this is the case for BTO in its ground-state rhombohedral structure (R3m, SG: 160). There is no symmetry breaking-induced intrinsic energy reduction on increasing the cell size.

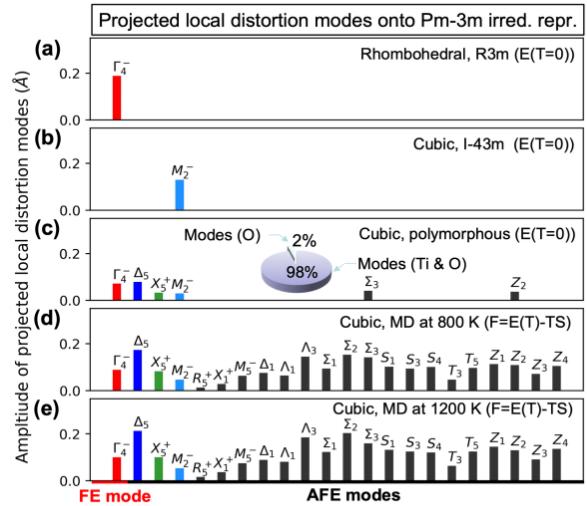

**Figure 3.** Projected distortion modes related to Ti and oxygen intrinsic- & thermal- displacements onto parent undistorted cubic (Pm-3m) irreducible representations of (a) ground state rhombohedral structure, (b) *ordered* anti-ferroelectric (AFE) cubic phase (I-43m), (c) polymorphous structure with intrinsic distortions, and (d, e) snapshots on equilibrated MD trajectory. Insert to c shows the indicated modes projected onto atomic displacements. The heights of bars in (c, d, and e) refer to the averaged amplitude for each of the 10, 100, and 100 configurations. The 10 configurations in (c) are obtained from 10 independent runs starting from the re-nudged polymorphous structure. Each of 100 configurations in (d, e) are extracted with interval 0.1 ps from the equilibrated MD trajectory with NPT ensemble at 800 and 1200 K, pressure 1 atm, and timestep 1 fs. $\Gamma_4^-$ is a ferroelectric (FE)-like Ti displacements, whereas the rest of the mode are AFE-like relative Ti displacements.

*The decomposition of intrinsic and dynamic distortion fields in terms of irreducible representations of the cubic Pm-3m symmetries (Fig. 3):* The polyDFT calculations show a local symmetry-breaking in cubic BTO although this calculation step excludes any effects due to the finite temperature. This points to an intrinsic energetic driving force for forming local distortions under the set of constraints given. It is interesting to inquire what is the *origin* of the intrinsic symmetry-breaking mode symmetries in the PE phase. We find that these symmetries are inherited from the low-temperature long-range ordered ferroelectric phase as well as theoretically predicted long-range ordered cubic anti-ferroelectric phase with net total dipole, [41] albeit without the long-range dipolar order. The low-temperature ferroelectric rhombohedral phase is a condensation of just one of these modes, the one associated with the $\Gamma_4^-$ irrep (Fig. 3a), whereas the predicted AFE phase is associated with $M_2^-$ irrep. We find that in the intrinsic polymorphous model of the PE phase, the projected $\Gamma_4^-$ irrep mode onto Pm-3m structure is still one of the most prominent modes (Fig. 3c), resulting in a Ti off-centering that is almost as large (0.09 Å) as in the rhombohedral phase (0.20 Å). However, the $\Gamma_4^-$ is not the only significant mode that is populated in the polymorphous network. Five

additional modes — $X_5^+$, $\Sigma_3$, $M_2^-$, $Z_2$, and $\Delta_5$ — all with AFE patterns have significant amplitudes. These modes originate from relative Ti displacements along opposite {100} directions in different TiO$_2$ layers. For example, for $X_5^+$, the AFE-like Ti displacements occur in adjacent TiO$_2$ layers, whereas for $\Delta_5$, the AFE-like Ti displacements occur in three TiO$_2$ layers, in which the Ti displacements in the middle TiO$_2$ layer are not displaced. These additional $X_5^+$, $\Sigma_3$, $M_2^-$, $Z_2$, and $\Delta_5$ modes have distortion magnitudes in the range of 0.02-0.08 Å. Importantly, they robustly appear in all independently restarted polymorphous calculations (Table S-1), suggesting they are a robust part of the polymorphous state and not due to incomplete sampling. Since $\Delta_5$, $\Sigma_3$, $M_2^-$, $Z_2$, and $X_5^+$ are absent in the globally rhombohedral phase, presumably, they play a role in relaxing the local distortion back towards the globally cubic unit cell.

*Thermal effects on paraelectric modes and Ti distributions:* To capture thermal effects, we use MD simulations (described in Supplementary computational details) starting from the intrinsic polymorphous results. The MD runs result in atomic configurations that can also be decomposed onto the basis of the distortional modes of irreps as was done for the polyDFT relaxed structures in Fig. 3c. As expected, as the temperature is raised, many additional modes become significantly populated. For example, the mode population averaged over 100 snapshots (within 10 ps timescale) in MD simulation at 800 and 1200 K are shown in Fig. 3(d, e), indicating that the $\Gamma_4^-$, $M_2^-$, and $X_5^+$ symmetry modes remain active even at 1200 K, i.e., well into the PE phase, despite the fact that new modes join in the symmetry breaking and the relative magnitude of each mode is temperature dependent. Further, the projected Ti displacements at 800 K and 1200 K (Fig. S-3) also indicate that the off-center distortion, though prominently off-center contribution at 800 K seems to be gradually overcome as temperature increase to 1200 K, being close center-flat-like distribution. The center-flat-like distribution has been observed based on the MD simulation with a force field and interpreted as mixed character of "order-disorder" and "displacive" [15,16] near the Curie temperatures. We interpret that the on-center population is overestimated by projections, which is strongly dependent on the display (Fig. S-1). Indeed, Pasiak et al. [16] found that the on-center population of Ti atoms is shallow at $T_c$ + 100K. Therefore, the vector of off-center might be changed due to thermal effect, but never be dominated population at zero as in cubic polymorphous work at T=0 K.

*The net polarity in the paraelectric phase and its dependence on temperature (Fig. 3):* The projections described above give important insights into the nature of the collective distortion modes of the system, but it gives incomplete information about the polarizability of the material. To quantify this, we need to consider the size of the local electric dipoles $\{\mu^{(i)}_{local}\}$, which are proportional to the displacements of Ti atoms away from the center of their local $O_6$ octahedral hosts. These Ti displacements can be represented with respect to the mass center of the octahedron in polymorphous structures and MD snapshots. The projections of Ti displacements along <100> direction is shown in Fig. 1(e, f) for 64 fu supercell for different temperatures, showing the presence of Ti-off-centering. This effect has been ascribed previously to the thermal effect,[44] but it emerges here from the static calculation, suggesting the origin is electronic symmetry-breaking associated with this energy lowering. The ensuing intrinsic net polarization is ~0.07 ± 0.01 Å oriented close to <111> direction before temperature set in. Further, by averaging all dipoles over a ~40 ps period on the equilibrium MD trajectories, we find that the global dipole is diminished with temperature: $\mu_{global}$ = 0.02 and 0.01 Å at 800 and 1200 K, respectively. This nonzero global polarization is due to the $\Gamma_4^-$ mode (Fig. 3) with a locally FE-like pattern. In our calculation, the net global polarization around the Curie temperature (Fig. S-4) is not zero, in agreement with the measured polarization of single crystal BTO above $T_c$. [44,45] This nonzero polarization might contribute to anomalous observations, such as Raman splitting, [46] SHG, [18] piezoelectricity [19,20] and seen in PE cubic BTO. As shown in Fig. 1(e, f) and S-3, the projected images show Ti displacements change off-center to on-center vibration from 800 to 1500 K (close to the melting point), which also depends on the definition of octahedral center and display plots (Fig. S-3). The off-center Ti displacements are observed at 1200 K (Fig. S-3) that might be overestimated because of overestimated lattice expansion by PBE functional. This finding is different from the general understanding (i.e., above the Curie temperature, the PE phase is described by the displacive model [47] as a single parabolic well, lacking static displacements) of paraelectric phase at finite temperatures.

We conclude that in addition to the usual thermally-driven disorder in conventional compounds such as $BaZrO_3$, other compounds such as PE $BaTiO_3$ manifest an *intrinsic* a-thermal precursor to symmetry-breaking in the form of a ('polymorphous') distribution of energy-lowering local motifs. These can survive at finite temperatures in the PE phase, providing a physical understanding of the microscopic structures and their anomalous phenomenon in these kinds of PE materials.

**Acknowledgements:**

We thank fruitful discussions with Prof. Ekhard K. H. Salje and Prof. Annette Bussmann-Holder on paraelectric, Prof. Zhi Wang on the long-range-order model of AFE, and Dr. Lin Ding Yuan for assistance in calculating polymorphous structures with different initial nudges shown in supplementary Table S-1. The work in the Zunger's group and Billinge's group was supported by the US National Science Foundation through grant DMREF-1921949. The calculations were done using the Extreme Science and Engineering Discovery Environment (XSEDE), which is supported by the National Science Foundation grant number ACI-1548562.